\documentclass[notoc,12pt,letterpaper,nohyper]{JHEP3}

\usepackage{epsfig,multicol,bbm,cite}

\def\be{\begin{equation}}
\def\ee{\end{equation}}
\def\bea{\begin{eqnarray}}
\def\eea{\end{eqnarray}}

\def\bbuildrel#1_#2^#3{\mathrel{\mathop{\kern 0pt#1}\limits_{#2}^{#3}}}
\def\slash#1{\setbox0=\hbox{$#1$}#1\hskip-\wd0\dimen0=5pt\advance
       \dimen0 by-\ht0\advance\dimen0 by\dp0\lower0.5\dimen0\hbox
         to\wd0{\hss\sl/\/\hss}}
\def\gev{{\rm GeV}}
\def\mev{{\rm MeV}}

\newcommand{\gae}{\lower 2pt \hbox{$\, \buildrel {\scriptstyle >}\over {\scriptstyle
\sim}\,$}}
\newcommand{\lae}{\lower 2pt \hbox{$\, \buildrel {\scriptstyle <}\over {\scriptstyle
\sim}\,$}}

\title{Hints for the scale of new CP-violating physics from B--CP anomalies}

\author{
Enrico Lunghi \\
Physics Department, Indiana University, Bloomington, IN 47405, USA \\
E-mail: \email{elunghi@indiana.edu}
}

\author{
Amarjit Soni \\
Physics Department, Brookhaven National Laboratory, Upton, NY 11973, USA \\
E-mail: \email{soni@bnl.gov}
}

\preprint{IUHET-525}

\abstract{
We consider several hints for new physics involving CP-asymmetries in B-decays and interpret them in terms of generic contributions to effective Wilson coefficients. The effects we focus on are: the differences in the fitted value of $\sin 2 \beta$ versus the ones directly measured via the time dependent CP asymmetries in $B \to J/\psi K$ or via $B\to (\phi,\eta^\prime) K$; the difference between the direct CP asymmetries in $B^- \to K^- \pi^0$ and $\bar B^0 \to K^- \pi^+$ and the $\approx 2.2 \sigma$ indications for the CP-asymmetry in $B_s \to J/\psi \phi$. To alleviate concerns regarding the disagreement between inclusive and exclusive $V_{ub}$, we show that our results hold even without the inclusion of $V_{ub}$ in the analysis. We find that no matter what kind of new physics (NP) is invoked to explain these effects, its effective scale is bounded from above from a few hundred GeV to a few TeV depending on specific assumptions regarding the type of new physics. The only exception to this is when the  NP contribution is assumed to reside entirely in  LR operators in K mixing, then the scale of NP can be as high as around 24 TeV; however, this case cannot account for CP asymmetry in $B_s \to J/\psi \phi$ or a difference in $\sin 2 \beta$ from penguin modes compared to that from $J/\psi K$ or for that matter  the large difference seen between direct CP asymmetries in $K^- \pi^+$ and in $K^- \pi^0$.
}

\begin{document}

\section{Introduction}
The only source of flavor changing interactions in the Standard Model (SM) is provided by the Cabibbo-Kobayashi-Maskawa (CKM) mixing matrix. Unitarity imposes testable constraints on the magnitude and phases of its elements; in particular, the relation
\bea
V_{td}^{} V_{tb}^* + V_{cd}^{} V_{cb}^* + V_{ud}^{} V_{ub}^* = 0
\label{CKMunitarity}
\eea
has received considerable attention because of its strong sensitivity to the single CP violating phase that appears in the CKM. The extraction of magnitudes and phases of the various terms in Eq.~(\ref{CKMunitarity}) from a large number of  observables is complicated by the presence of hadronic uncertainties. Due to the superb performance of the two asymmetric B-factories, it has been established that the Standard Model's CKM-paradigm~\cite{ckm} works to an accuracy of around ~15-20\%~\cite{Browder:2008em}; therefore the effects of New Physics (NP) are expected to be a perturbation and sub-dominant. The search for NP therefore requires very good control over theory errors and high statistics data from experiments. The combination of very precise experimental results from the B--factories BaBar and Belle and from the Tevatron experiments CDF and D0, with recent progress in lattice QCD, namely the improved determination of $B_K$~\cite{Antonio:2007pb} and of the $B\to \pi$ form factor~\cite{Ruth08}, leads to the emergence of several indications of possible deviations from the SM. We focus on the following issues:

{\bf \em (a)} The deduced value of $\sin 2\beta$\footnote[2]{We recall the usuage of two equivalent notations: $(\phi_1,\phi_2,\phi_3)\equiv (\beta, \alpha, \gamma)$.} differs from the directly measured value at the $2\sigma$ level. The observables that are used to deduce the value of $\sin 2 \beta$ are: the determinations of $|V_{ub}|$ and $|V_{cb}|$ from inclusive and exclusive $b\to (u,c) \ell \nu$ decays, the (indirect) CP violating parameter $\varepsilon_K$, and the ratio of the meson--antimeson mass differences in the $B_s$ and $B_d$ systems. In this deduction of $\sin 2 \beta$ the only CP-violating quantity is that in the neutral kaon system. If the SM description of CP-violation through the CKM-paradigm with a single CP-odd phase is correct, then the value of $\sin 2 \beta$ thus obtained should agree with the  directly measured value of $\sin 2 \beta$ in B-factory experiments. Following this logic, we will term this deduced value of $\sin 2 \beta$ as $\sin 2 \beta_{\rm SM}$. Actually, since there is some $\sim 2 \sigma$ discrepancy between the values of $|V_{ub}|$ extracted from inclusive and inclusive decays~\cite{Yao:2006px}, there is considerable motivation for obtaining $\sin 2 \beta_{\rm SM}$ without using $|V_{ub}|$ which we call $\sin 2 \beta_{\rm SM}^{\rm no V_{ub}}$. Of course $\sin 2\beta_{\rm SM}^{\rm no V_{ub}}$ will only be of use if it has reasonably small errors.  That this has become possible, due to the improved determination of $\hat B_K$ from the lattice, was recently empahsized in~\cite{Lunghi:2008aa}. 

There are two important ways for extracting directly $\sin 2 \beta$ via measurements of time dependent CP asymmetries. First there is the gold-plated ({\it i.e.} free from hadronic uncertainties to a very high degree of accuracy) measurement of $\sin 2 \beta$ via the time dependent CP asymmetry in $B\to (J/\psi K)$~\cite{bigi_sanda}; for clarity we will denote this as $\sin 2 \beta_{\psi K}$. A second way to measure $\sin 2 \beta$ is via $b\to s$ penguin transitions such as $B$ decays to $\phi K_S$, $\eta' K_S$, $K_S K_S K_S$, $\pi^0 K_S$, $\rho^0 K_S$, $\omega K_S$, $f_0 K_S$, $\pi^0 \pi^0 K_S$, $\phi \pi^0 K_S$, $K^+ K^- K^0$~\cite{gw,london}, etc. Unfortunately, this method has some hadronic uncertainties. In the original papers this was crudely estimated at $\approx \lambda^2 \approx 5\%$~\cite{gwi,london}. In the past few years these modes received considerable theoretical attention~\cite{Beneke:2005pu,ccs2,ccs3,nir,jure,lu} and as a result of that it now seems that amongst the two body modes, the $\eta' K_s$ and $\phi K_s$ final states receive hadronic corrections at the few percent level and are therefore very clean. For this reason, in the present work we will only include these two penguin modes and we will term $\sin 2 \beta$ extracted from their weighted average as $\sin 2 \beta_{(\phi,\eta^\prime)K}$

If the CKM description of CP violation is correct then all three determinations of $\sin 2 \beta$ should agree with each other. In fact, both the ``predicted'' values of $\sin 2 \beta$, whether one uses $|V_{ub}|$ or not, differ from the directly measured values by $\sim 2 \sigma$.

{\bf \em (b)} New physics in $b\to s$ penguin amplitudes is also hinted at by the fact that $\sin 2 \beta_{\psi K}$ differs from $\sin 2 \beta_{(\phi,\eta^\prime)K}$ mentioned above by  around $1.5 \sigma$. 

In passing, we want to briefly mention that there is another feature of the time dependent CP-asymmetry measurements in various penguin modes\footnote[2]{Time dependent CP asymmetries in a generic $B\to f$ mode are denoted with the symbol $S_f$.} that deserves discussion. While the difference between  $S_{J/\psi K_S}$ and $S_{\rm penguin}$, for each of the penguin modes is not that significant, a specially intriguing feature is that for many modes the central values of the asymmetry tends to be smaller than $S_{J/\psi K_S}$. Since $\sin 2 \beta_{\psi K}$ is less than $\sin 2 \beta_{SM}$, this obviously implies that the central value of $\sin 2 \beta_{\rm penguin}$ for almost all modes is also smaller than $\sin 2 \beta_{SM}$.

It is also useful to recall that comparison of $S_{J/\psi K_S}$ with $S_{(\phi,\eta^\prime)K}$ indicates whether or not $b\to s$ penguin transitions are affected by NP. Since the time dependent CP in {\it both} tree and penguin modes necessarily involves $B_d \bar B_d$ oscillations, comparison between $\sin 2 \beta_{\psi K}$ and $\sin 2 \beta{(\phi,\eta^\prime)K}$  does not tell us anything about whether there is new physics in $B_d$ mixing.

To the extent that $\sin 2 \beta_{\psi K}$ differs from $\sin 2 \beta_{SM}$, the possibilty of NP contributions to $B_d$ and/or $K$ mixing (emphasized in particular in Ref.~\cite{Buras:2008nn}) cannot be ruled out. Furthermore,  once NP is invoked, we need to be careful in identifying which observables are sensitive to the type of NP that may be out there.

{\bf \em (c)} Another hint that $b \to s$ penguin transitions may be exhibiting a non-standard CP-odd phase comes from the comparison of the partial rate asymmetry in $B^0 \to K^+ \pi^-$ and $B^+ \to K^+ \pi^0$. Experimentally this difference has been determined to be $14.4 \pm 2.9 \%$~\cite{Yao:2006px}. These two decays are closely related as they simply require switching the spectator (u,d) quarks. Therefore, the difference between these asymmetries vanishes in the limit of exact isospin and should be small. In sharp contrast, experimentally the two asymmetries are found to have an opposite sign and the result $14.4\%$ is non-vanishing by over 4 $\sigma$. It is difficult to rigorously assess the full significance of this unexpectedly large difference since we cannot reliably calculate, in a model independent fashion, the expectation from  the SM taking QCD fully into account.  In the QCD factorization approach~\cite{Beneke:2003zv,Beneke:2002jn}, the predictions for BR and CP asymmetries in hadronic two--body $B$ decays suffer from very large hadronic uncertainties, rendering problematic their use in NP searches. However, see for instance Refs.~\cite{Lunghi:2007ak}, a case can be made for NP in the difference $\Delta A_{CP} = A_{CP} (B^- \to K^- \pi^0) - A_{CP} (\bar B^0 \to K^- \pi^+)$. Ref.~\cite{Lunghi:2007ak,Cheng:2004ru} show that in the QCDF analysis of this quantity, most parametric uncertainties that occur for individual asymmetries cancel out and the theoretical prediction becomes  under reasonable control (see Ref.~\cite{Gronau:2006ha} for an alternate point of view) yielding $(2.2 \pm 2.4)\%$ which is still about $3.5 \sigma$ away from the measured value.

{\bf \em (d)} The time dependent CP asymmetry in $B_s \to J/\psi \phi$ is free from hadronic uncertainties, has been recently measured at CDF~\cite{Aaltonen:2007he} and D0~\cite{Abazov:2008fj,Ellison:2009sg} and deviates from the SM at the $2\sigma$ level~\cite{Lenz:2006hd,Barberio:2008fa}. Beyond the SM, a CP-odd phase in $B_s$ mixing is required in order to explain this discrepancy.

{\bf \em (e)} As is well known, over the past decade or so methods have been developed that allow direct measurement of all three angles of the unitarity triangle, ($\alpha$, $\beta$, $\gamma$)~\cite{Browder:2008em}. What makes these methods so atractive and useful is that attempts are made either not to use any theoretical input or assumptions or make the minimal use if necessary. For the angle $\beta$ (via time dependent CP studies of {\it e.g.} $B \to \psi K_s$~\cite{bigi_sanda}) and the angle $\gamma$ (via direct CP studies in $B^{\pm}\to (DK, D^* K, D K^*)$ and/or time dependent CP measurements in $B^0, \bar B^0 \to (DK, D^* K, D K^*)$~\cite{Gronau:1991dp,Gronau:1990ra,Atwood:1996ci,Atwood:2000ck,Atwood:2001ww,Giri:2003ty}) no theoretical assumptions are needed. The resulting precision is largely data driven with an irreducible theory error of $O(0.1\%)$ for $\beta$ and $ < 0.1\%$ for gamma~\cite{Browder:2008em}. For extracting the angle $\alpha$ a simple time dependent study of $B^0, \bar B^0 \to \pi^+ \pi^-$ does not suffice and an isospin analysis~\cite{Gronau:1990ka,Snyder:1993mx,Quinn:2000by} becomes necessary entailing a somewhat larger irreducible theory error, O(few \%). Currently, the angle $\alpha$ is being extracted by using branching ratios and CP asymmetries in $B\to (\pi\pi, \rho\rho, \rho\pi)$.

Thus another useful avenue to exploit in order to test the CKM-paradigm and to constrain NP is to fit the Unitarity Triangle utilizing only the three angles ($\alpha$, $\beta$, $\gamma$) which are directly measured without theoretical assumptions or input. We will use this approach to extract the resulting values of the Wolfenstein parameters $\rho$ and $\eta$ and compare them with the values deduced from the use of other methods and inputs. In particular, we show that once again the use of appropriate lattice matrix elements for $\epsilon_K$, $\Delta M_{s,d}$, $V_{cb}$ and with or without $V_{ub}$ leads to $\sim 2 \sigma$ deviations in one or both of the Wolfenstein parameters.

{\bf \em (f)} Finally, we mention another approach to determine the fitted value of $\sin 2 \beta$ that does not use $V_{ub}$ determined from exclusive or inclusive semi-leptonic decays, rather this  makes use of the directly measured values of $\alpha$ and $\gamma$ (see point {\it (e)} above) along with $\epsilon_K$ and the ratio of mass differences in $B_s$ and $B_d$ mesons (see point {\it (a)} above). In the tables in Fig.~\ref{fig:fitresults} we list different ways of arriving at the fitted values of $\sin 2 \beta$ and indicate the resulting tensions. 

These anomalies in CP asymmetries involving $B$ and $B_s$ mesons that we mention above are all at the $(2\div 3) \sigma$ level and may be indicative of new physics in $B_d$, $B_s$ and/or in $K$ mixing and also in $b \to s$ penguin transitions. Note that  this new physics necessarily has to carry with it a beyond the SM CP-odd phase as all the observables being discussed here involve CP violation. 

In this paper we analyze these  hints for new physics from an effective theory point of view: we parametrize NP contributions to various operators in terms of effective scales that, with the aid of these measurements, turn out to be constrained from above and from below. The interpretation of the two sigma discrepancies described above in terms of upper bounds on the scale of new physics has a caveat. In our approach NP contributions are essentially proportional to $1/\Lambda^2$ where $\Lambda$ is a generic NP scale. The tensions that we discuss translate into $1/\Lambda^2 > 0$ (and hence $\Lambda < \infty$) at the $2\sigma$ level; therefore, we are able to set an upper limit only at the 68\% (and maybe 95\%) C.L.. Beyond this confidence the measurements we are considering are compatible with the SM and no upper bound on NP is implied.

The paper is organized as follows. In Sec.~\ref{sec:sm} we summarize the present state of the various fits that we use. In Sec~\ref{sec:mia} we perfom a model independent analysis of new physics in $B_d /K$ mixing and in $b\to s$ penguin amplitudes. In Secs.~\ref{sec:oa-a} and \ref{sec:oa-b} we interpret the discrepancies in the fit to the unitarity triangle (UT) and in penguin amplitudes in terms of NP contributions to various operators. In Sec.~\ref{sec:discussion} we summarize and discuss our findings. 

\section{Present status of the SM fits}
\label{sec:sm}
The set of inputs that we use in the fit is summarized in Table~\ref{tab:utinputs}. $\alpha_{\pi\pi,\rho\rho,\rho\pi}$ and $\alpha_{\rho\rho}$ are the direct determinations of $\alpha$ that we obtain from the isospin analysis of $B\to (\pi\pi,\rho\rho,\rho\pi)$ and $B\to \rho\rho$ decays, respectively (we use the latter when discussing for NP effects in mixing so as to avoid pollution from possible NP contributions from $b\to d$ penguins). The direct determination of $\gamma$ is taken from the model independent UTfit analysis of $B\to D^{(*)} K^{(*)}$ decays~\cite{Bona:2005vz,Bona:2006ah}.
\begin{table}[t]
\begin{center}
\begin{tabular}{ll}
\hline\hline
$\left| V_{cb} \right| = \cases{
(41.67 \pm 0.68) \times 10^{-3}  \;\; \hbox{incl~\cite{Barberio:2008fa}}   \cr
(38.7 \pm 1.35) \times 10^{-3} \;\; \hbox{excl~\cite{Laiho:2007pn}}  \cr 
(41.0 \pm 0.63) \times 10^{-3}\;\; {\rm comb} \cr
}$ &
$\left| V_{ub} \right| = \cases{
(39.6^{+2.5}_{-2.7}) \times 10^{-4} \;\; \hbox{incl~\cite{Barberio:2008fa}}  \cr
(33.8 \pm 3.5) \times 10^{-4} \;\; \hbox{excl~\cite{Ruth08}} \cr 
(37.4 \pm 2.1) \times 10^{-3}\;\; {\rm comb} \cr}$ \vphantom{$\cases{a & b \cr a & b \cr a & b \cr &  \cr}$} \\
$\Delta m_{B_d} = (0.507 \pm 0.005)\; {\rm ps}^{-1}$~\cite{Barberio:2008fa}  & 
$\Delta m_{B_s} = (17.77 \pm 0.10 \pm 0.07)\;  {\rm ps}^{-1}$~\cite{Evans:2007hq}  \vphantom{\Bigg(} \\
$\Delta A_{CP} = ( 14.8 \pm 2.8 ) \%$~\cite{Barberio:2008fa} & 
$\gamma = (78 \pm 12)^{\rm o}$~\cite{Bona:2005vz,Bona:2006ah}
 \vphantom{\Big(} \\
$\alpha_{\pi\pi,\rho\rho,\rho\pi} = (88.7 \pm 4.7)^{\rm o}$ &
$\alpha_{\rho\rho} = (87.8 \pm 5.6)^{\rm o}$ 
\vphantom{\Big(}  \\
$\eta_1 = 1.51 \pm 0.24$~\cite{Herrlich:1993yv} & 
$m_{t, pole} = (172.4 \pm 1.2) \; {\rm GeV}$~\cite{:2008vn} \vphantom{\Big(}\\
$\eta_2 = 0.5765 \pm 0.0065$~\cite{Buras:1990fn}  & 
$m_c(m_c) = (1.224 \pm 0.057 ) \; {\rm GeV}$~\cite{Hoang:2005zw}\vphantom{\Big(}\\
$\eta_3 = 0.47 \pm 0.04$~\cite{Herrlich:1995hh}  &  
$\varepsilon_K = (2.232 \pm 0.007 ) \times 10^{-3}$  \vphantom{\Big(} \vphantom{\Big(}\\
$\eta_B = 0.551 \pm 0.007$~\cite{Buchalla:1996vs} &
$S_{J/\psi \phi} = \left( -0.76^{+0.37}_{-0.33} \; \vee \; -2.37^{+0.33}_{-0.37} \right)$~\cite{Ellison:2009sg}  \vphantom{\Big(}  \\ 
$S_{\psi K_S} = 0.672 \pm 0.024$~\cite{Barberio:2008fa} &
$S_{\phi K_S} = 0.44^{+0.17}_{-0.18}$~\cite{Barberio:2008fa}   \vphantom{\Big(} \\
$S_{\eta^\prime K_S} = 0.59 \pm 0.07$~\cite{Barberio:2008fa} &
$f_{B_s} \sqrt{\hat B_{B_s}} = (0.304 \pm 0.032 )\; \gev$~\cite{Tantalo:2007ai} \vphantom{\Big(}\\
$\xi  = 1.211 \pm 0.045$~\cite{Evans:2008} & 
$\kappa_\varepsilon = 0.92 \pm 0.02$~\cite{Buras:2008nn}  \vphantom{\Big(} \\
$\hat B_K = 0.720 \pm 0.013 \pm 0.037$~\cite{Antonio:2007pb}   &
$f_K = (155.5 \pm 0.2 \pm 0.8 \pm 0.2) \; \mev$~\cite{Barberio:2008fa}
  \vphantom{\Big(} \\
  $\lambda = 0.2255  \pm 0.0007$~\cite{Antonelli:2008jg} &  \vphantom{\Big(} \\
\hline
\hline
\end{tabular}
\caption{Inputs used in the unitarity triangle fit. Quantities not explicitly given are taken from Ref.~\cite{Yao:2006px}.\label{tab:utinputs}}
\end{center}
\end{table}
The explicit expressions for $\Delta M_{B_q}$, $\Delta M_{B_s}/ \Delta M_{B_d}$ and $\varepsilon_K$ in the SM are:
\bea
\Delta M_{B_q} & = &  
 2 \left| M_{12}^q \right| =  \frac{\left| \langle \bar B_q^0 | {\cal H}_{\rm eff} | B_q^0 \rangle  \right|}{ m_{B_q}} = 
\frac{G_F^2}{12  \pi^2 }  m_W^2 m_{B_q}  f_{B_q}^2  \hat B_{B_q} \eta_B S_0 ( x_t)  \left| V_{tb}^{} V_{tq}^* \right|^2 
\; , \\
\frac{\Delta M_{B_s}}{\Delta M_{B_d}} & = & 
\xi^2 \frac{m_{B_s}}{m_{B_d}} \left| \frac{V_{ts}}{V_{td}} \right|^2 \; ,\\
| \varepsilon_K | & = & \frac{G_F^2 m_W^2 f_K^2 m_K}{12 \sqrt{2} \pi^2 \Delta m_K^{\rm exp}} 
 \hat B_K  \kappa_\varepsilon \; {\rm Im} \Big(
\eta_1 S_0 (x_c) \left( V_{cs}^{} V_{cd}^* \right)^2 
+2 \eta_3 S_0 (x_c,x_t) V_{cs}^{} V_{cd}^*  V_{ts}^{} V_{td}^* \nonumber \\
& & +\eta_2 S_0 (x_t) \left( V_{ts}^{} V_{td}^* \right)^2 
\Big) \; . \label{ek}
\eea
The $B$ parameters (for $M=K, \; B_d , \; B_s$) parametrize the matrix elements
\bea
 \langle \bar M | Q_1(\mu) | M \rangle & = & \frac{2}{3}m_M^2 f_M^2 B_M^{\overline{\rm MS}} (\mu) 
\eea
\FIGURE[t]{
\epsfig{file=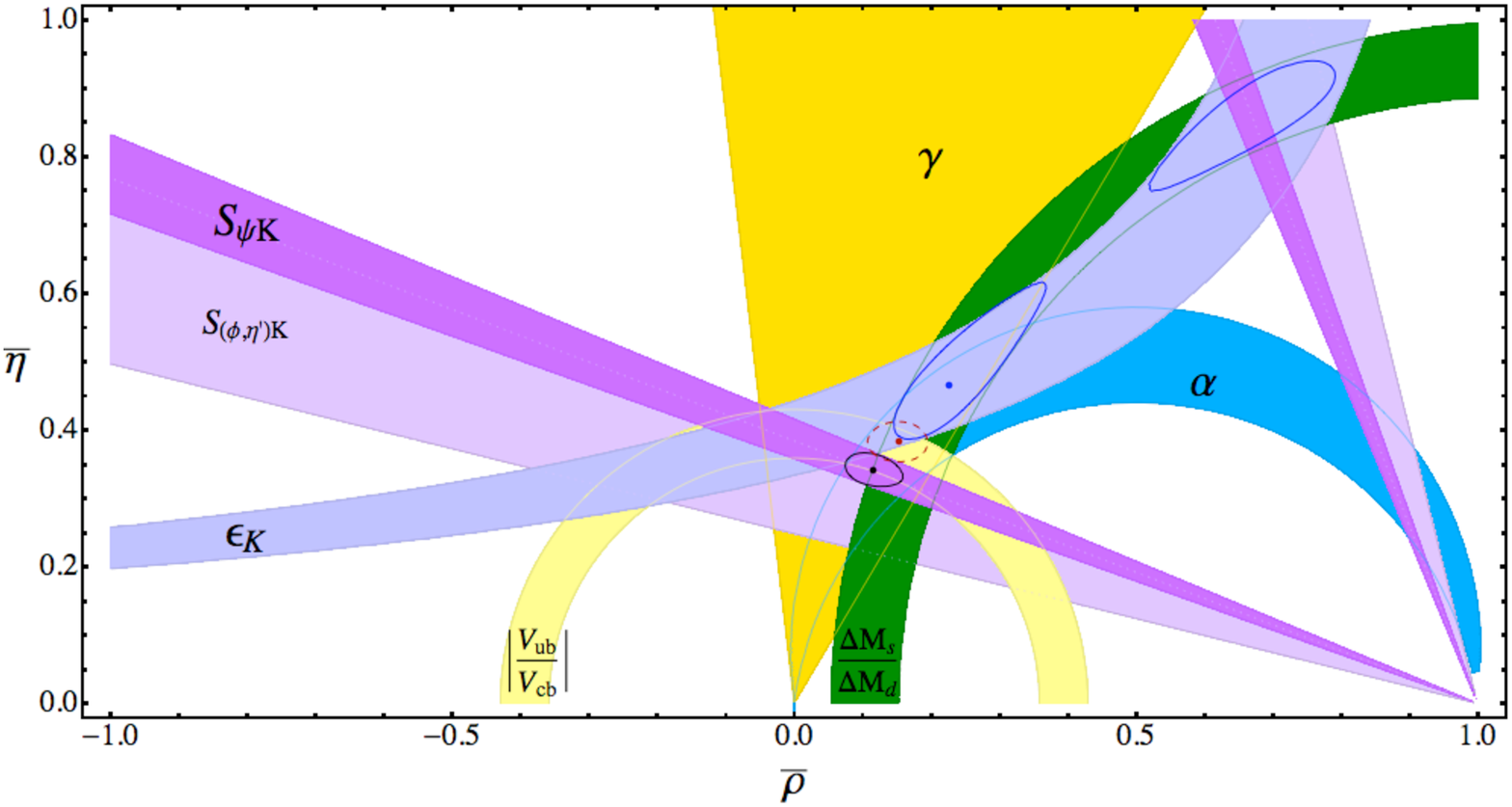,width= 0.7\linewidth}
\caption{Unitarity triangle fit in the SM (68\% C.L.). The solid black, solid blue and dashed red contours are obtained using ($\alpha$, $\beta$, $\gamma$), ($\varepsilon_K$, $\Delta M_{B_s}$, $\Delta M_{B_d}$, $V_{cb}$) and ($\varepsilon_K$, $\Delta M_{B_s}$, $\Delta M_{B_d}$, $V_{cb}$, $V_{ub}$), respectively. \label{fig:utsm}}}
where the operator $Q_1$ is given in Eq.~(\ref{q1}). The $\hat B_M$ parameters are renormalization group invariant quantities and differ from the corresponding $B_M^{\overline{\rm MS}} (\mu)$ by a perturbative factor (see, for instance, Refs.~\cite{Buchalla:1996vs,Buras:1998raa} for the details of this standard procedure). The quantity $\kappa_\varepsilon$ comes from the inclusion in $\varepsilon_K$ of the $I=0$ component of the $K\to \pi \pi$ amplitude~\cite{Buras:2008nn,Buras:2009pj,Anikeev:2001rk,Andriyash:2003ym,Andriyash:2005ax}. The loop--functions can  be found, for instance, in Ref.~\cite{Buras:1998raa, Buras:2008nn}. In this paper we will not concern ourselves with possible NP contributions to EW operators in the Kaon sector, whose effect is to alter the extraction of the factor $\kappa_\varepsilon$ from data on $\varepsilon^\prime/\varepsilon$ (See Ref.~\cite{Buras:2009pj} for a complete discussion of this issue).

\begin{figure}[t]
\begin{center}
\begin{tabular}{rcccc} \hline
& $\bar \rho$ & $\bar\eta$ & $\sin 2\beta$ \cr
$\varepsilon_K, \Delta M_{B_{d,s}}, V_{cb}$: & 
$0.236 \pm 0.063$ & $0.478 \pm 0.066$ & $0.885 \pm 0.082$
\cr
$\varepsilon_K, \Delta M_{B_{d,s}}, V_{cb},V_{ub}$: & 
$0.152 \pm 0.028$ & $0.383 \pm 0.019$ & $0.749 \pm 0.030$
\cr
$\varepsilon_K, \Delta M_{B_{d,s}}, V_{cb},\alpha,\gamma$: & 
$0.209 \pm 0.049$ & $0.442 \pm 0.046$ & $0.846 \pm 0.069$
\cr
$\varepsilon_K, \Delta M_{B_{d,s}}, V_{cb},\alpha,\gamma,V_{ub}$: & 
$0.150 \pm 0.024$ & $0.382 \pm 0.018$ &  $0.747 \pm 0.029$
\cr
$\alpha,\gamma,S_{\psi K}$: & 
$0.116 \pm 0.027$ & $0.341 \pm 0.016$ &
\vphantom{$\Big($} 
\cr
$\varepsilon_K, \Delta M_{B_{d,s}}, V_{cb},V_{ub},\alpha,\gamma,S_{\psi K}$: & 
$0.127 \pm 0.020$ & $0.357 \pm 0.012$ &
\vphantom{$\Big($} 
\cr \hline
\end{tabular}
\epsfig{file=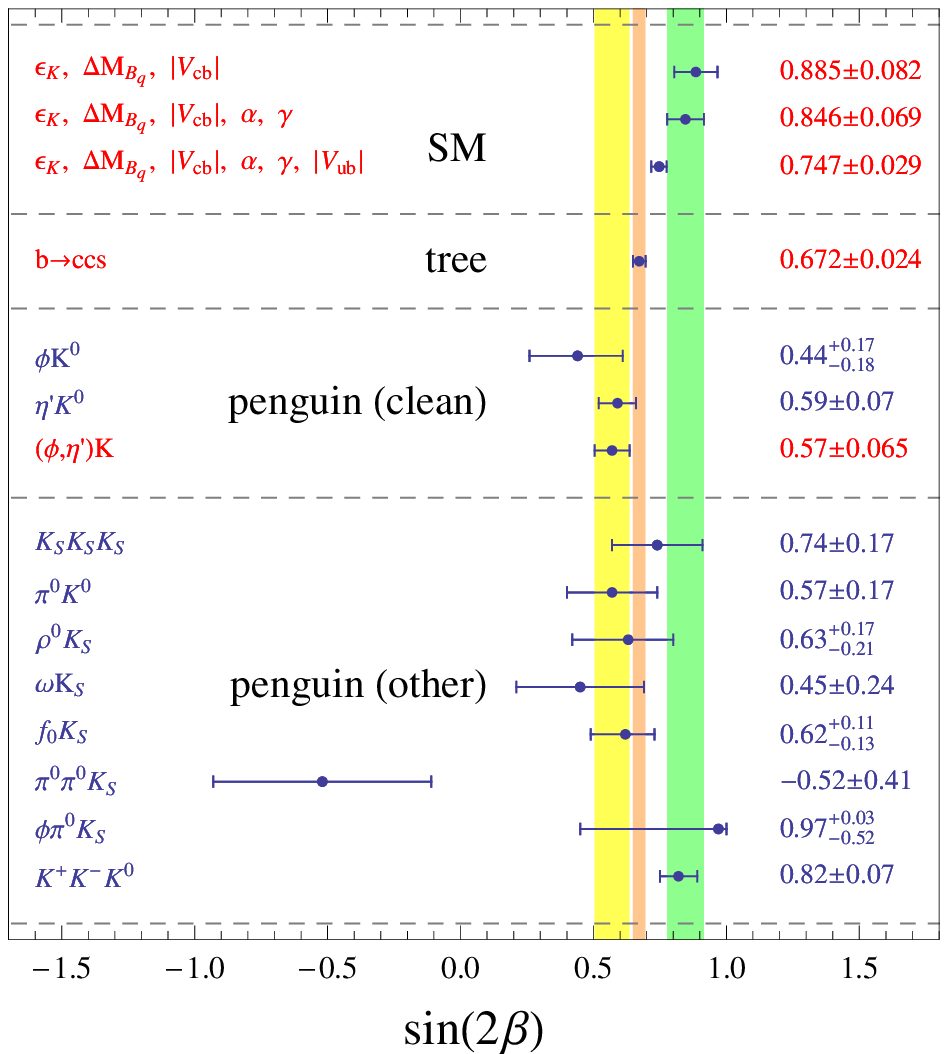,width= 0.5\linewidth}
\raisebox{4.5cm}{
\begin{tabular}{c|ccc} 
mode  & w/out $V_{ub}$ &  with $V_{ub}$ \\ \hline
$S_{\psi K_S}$ &  $ 2.4 \; \sigma$ & $ 2.0 \; \sigma$ \\ 
$S_{\phi K_S}$ &  $ 2.2 \; \sigma$ & $ 1.8 \; \sigma$ \\  
$S_{\eta^\prime K_S}$& $ 2.6 \; \sigma$ & $ 2.1 \; \sigma$ \\  
$S_{(\phi+\eta^\prime) K_S}$ & $ 2.9 \; \sigma$ & $ 2.5 \; \sigma$ \\  
\end{tabular}
}
\caption{Results of the fit to the unitarity triangle within the SM. In the table on top we collect the results we obtain for different selection of inputs. The plot is a graphical comparison between the SM predictions given above and the direct determinations in $b\to c\bar c s$ and $b\to s$ penguin modes. In bottom-right table we show the deviation of the clean $\sin 2\beta$ measurements from the SM predictions obtained using $\varepsilon_K$, $\Delta M_{B_q}$, $|V_{cb}$, $\alpha$, $\gamma$. The last column shows the impact of $|V_{ub}|$.
\label{fig:fitresults}}
\end{center}
\end{figure}

Our fitting procedure consists in writing a chi-squared that includes all experimental measurements and lattice determinations. This procedure implies that all systematic errors are treated as gaussian. While the true nature of systematic uncertainties remains subject of debate (see, for instance, the prescriptions adopted in Refs.~\cite{Ciuchini:2000de,Charles:2004jd}), we believe that our choice is preferable to flat systematic pdf's for several reasons: gaussian systematics lead to more conservative determinations of confidence level intervals; moreover, systematic errors in both lattice QCD and experiments are usually obtained by combining multiple sources of uncertainties, thus partially justifying our assumption. As usual we adopt the Wolfenstein parametrization of the CKM matrix and truncate the expansion at $O(\lambda^4)$. The $68\% C.L.$ allowed regions in the $(\bar\rho,\bar\eta)$ plane are shown in Fig.~\ref{fig:utsm}. In Fig ~\ref{fig:fitresults} we summarize the numerical results that we obtain for $\bar\rho$, $\bar\eta$ and $\sin 2 \beta$. In order to better illustrate the anatomy of the $2\sigma$ tension we decided to present the results corresponding to the inclusion of different sets of observables in the 
fit (we show also the result of the complete fit for comparison). It is interesting to note that the model independent determinations of $\alpha$ and $\gamma$ affect the fit only in absence of the $|V_{ub}|$ constraint. In the figure we also include a graphical representation of the discrepancy between direct and indirect (SM prediction) determinations of $\sin 2\beta$ as well as a pull table in which we quantify this discrepancy in terms of standard deviations. In the pull table, the {\it w/out} ({\it with}) $V_{ub}$ column refers to the treatment of $V_{ub}$ on top of  a fit that includes $\varepsilon_K$, $\Delta M_{B_q}$, $|V_{cb}|$, $\alpha$ and $\gamma$; the reference values of the SM predictions that we use in the pull table are therefore:
\bea
\sin(2\beta) = \cases{
0.846 \pm 0.069 & w/out $|V_{ub}|$ \cr
0.747 \pm 0.029 & with  $|V_{ub}|$  \cr
} .
\label{sin2betaprediction}
\eea
These results summarize nicely the $2\sigma$ tensions {\bf \em (a)} and {\bf \em (b)} that we discussed in the introduction. 

\section{Model independent analysis}
\label{sec:mia}
The results of the previous analysis can be interpreted in the context of new physics contributions to $B_d$-mixing ($M_{12}^{B_d}$) , $\varepsilon_K$ and to $b\to s$ penguin amplitudes ($A_{b\to s}$). For the sake of simplicity we consider only the two extreme scenarios in which we admit new physics effects to $(M_{12}^{B_d}, A_{b\to s})$ and $(\varepsilon_K, A_{b\to s})$, respectively. In this section we adopt very general parametrizations of possible new physics contributions; the connection to actual mass scales will be discussed in Secs.~\ref{sec:oa-a} and \ref{sec:oa-b}.

\subsection{Scenario I}
\label{subsec:mia1}
We assume all new physics effects to be effectively taken into account via the introductions of two extra phases, $\phi_d$ and $\theta_A$: 
\bea
M_{12}^{d} & = & \left( M_{12}^{d} \right)_{\rm SM} \; e^{2 i \phi_d} \; , \label{m12}\\
{\cal A}_{b\to  s} & = &\left( {\cal A}_{b\to s} \right)_{\rm SM} \; e^{ i \theta_A} \;. \label{abs}
\eea
The expressions for the time--dependent CP asymmetries become:
\bea
S_{\psi K} & = & \sin \left[ 2 (\beta + \phi_d) \right] \; , \\
S_{(\phi,\eta^\prime)K} & = & \sin \left[ 2 (\beta + \phi_d + \theta_A) \right] \; .
\eea
Note that we are implicitly assuming that NP effects in $b\to s$ penguin amplitudes are identical in the $\phi$ and $\eta^\prime$ modes. This is necessary in order to use a simple parametrization as in (\ref{abs}). This assumption will be relaxed in the operator level analysis presented in Sec.~\ref{sec:oa-b} where we adopt QCD factorization. Furthermore, the extraction of $\gamma$ from $B\to D^{(*)} K^{(*)}$ decays is controlled by tree-level decays and is assumed to be insensitive to new physics effects. This assumption does not hold for $\alpha$. In this case the isospin analysis extracts and effective angle given by $\alpha_{\rm eff} = \alpha - \phi_d + \theta_{\rm penguin}$. Here $\phi_d$ is the same angle that appears in Eq.~(\ref{m12}), while $\theta_{\rm penguin}$ is a possible new physics phase in $b\to d$ penguin amplitudes. In order to simplify the analysis we will utilize only the $B\to \rho\rho$ channels in the extraction of $\alpha_{\rm eff}$ because, in this case, the penguin contribution turns out to be experimentally very small and we are justified in setting $\theta_{\rm penguin} = 0$. In this scenario, the theory prediction for $\sin (2\beta)$ is obtained by excluding from the chi-squared both $S_{\psi K}$ {\it and} $\alpha$; we obtain:
\bea
\sin(2\beta) = \cases{
0.867 \pm 0.080 & without $|V_{ub}|$ \cr
0.747 \pm 0.029 & with  $|V_{ub}|$  \cr
} .
\eea
\FIGURE[t]{
\epsfig{file=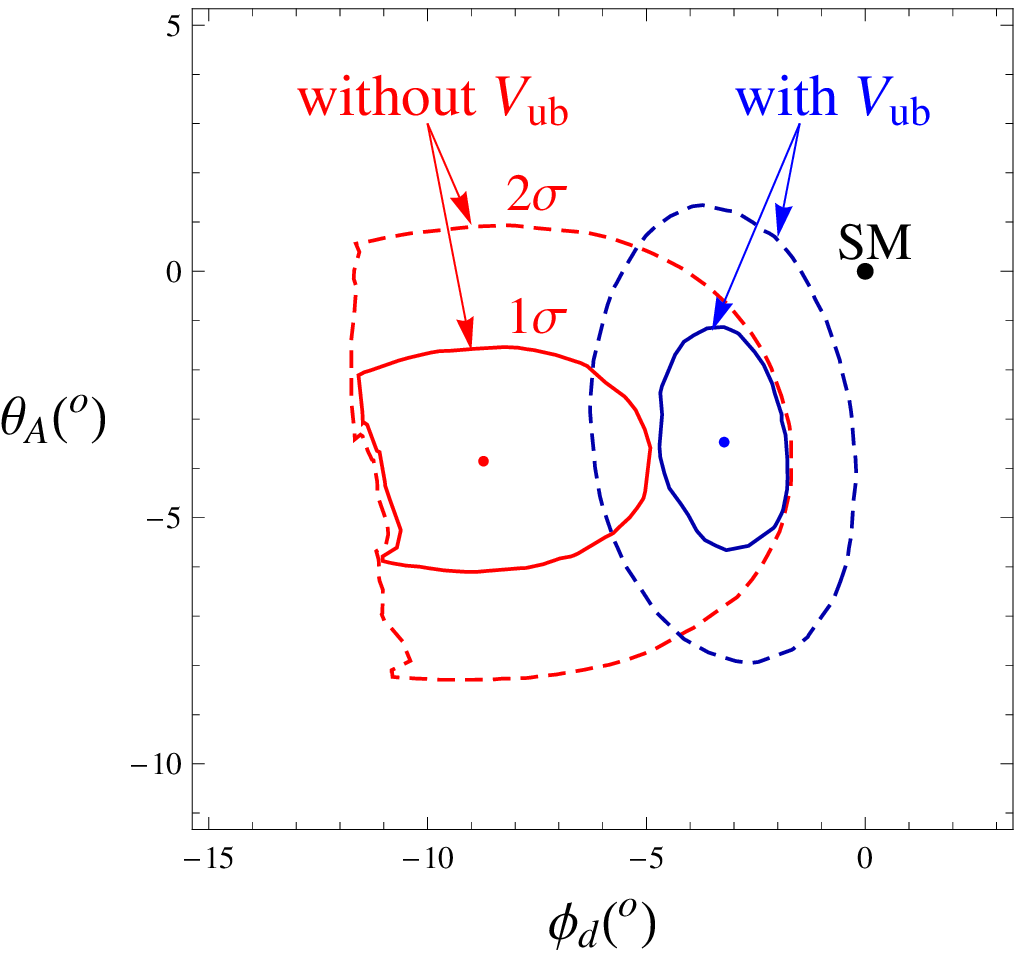,width= 0.45\linewidth}
\epsfig{file=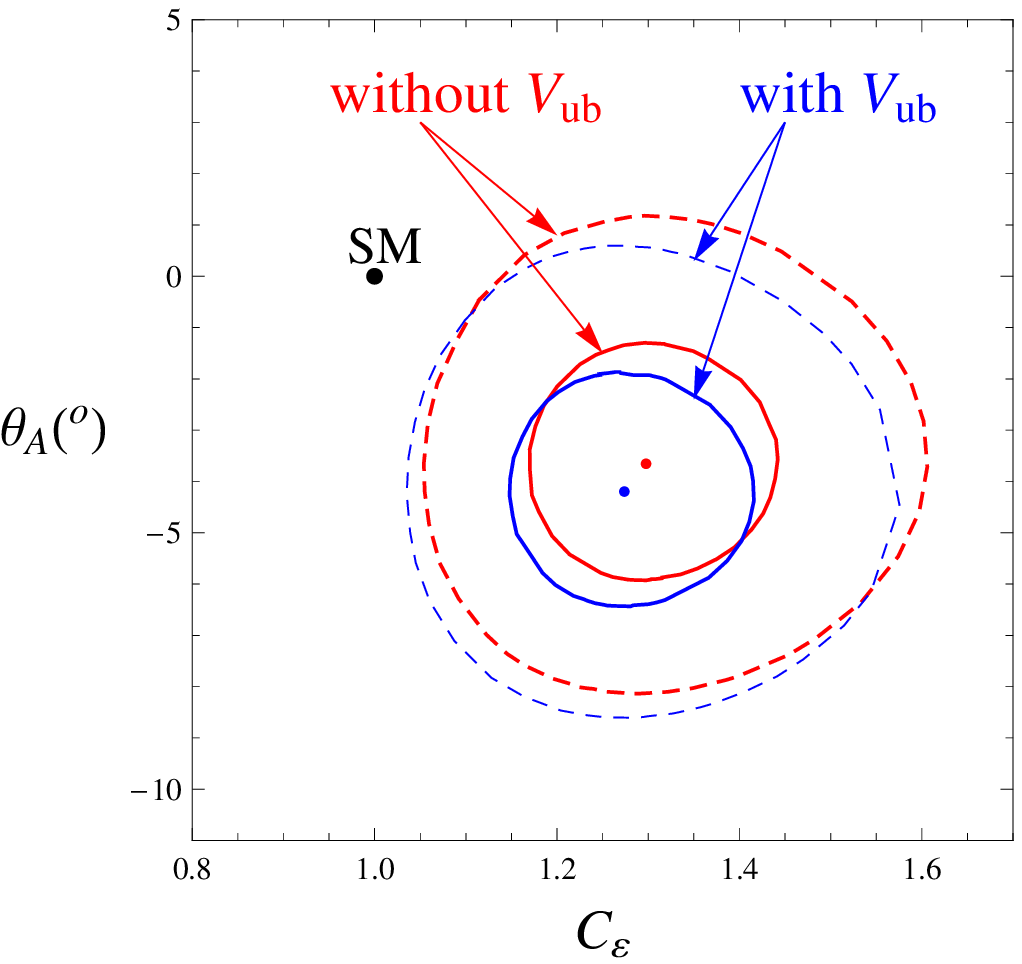,width= 0.45\linewidth}
\caption{
Model independent analysis of new physics effects in $B_d$ (left panel), $K$ (right panel) mixing and ${\cal A}_{b\to s}$ (both panels). Solid and dashed lines corresponds to the $1\sigma$ and $2\sigma$ contours, respectively. \label{fig:mia}}
}
In the left panel of Fig.~\ref{fig:mia} the contours define regions whose projections on the axes yield the one dimensional ranges at 68\% (solid) and 95\% (dashed) confidence level. Without the inclusion of $V_{ub}$ we obtain:
\bea
\phi_d &=& -\left( 8.7_{-2.8}^{+3.4} \right)^{\rm o} \label{mia1-phid-novub}\\ 
\theta_A &=& -\left( 3.8 \pm 2.1 \right)^{\rm o} \label{mia1-thetaA-novub}
\eea
As can be seen from Fig.~\ref{fig:mia} the negative error on $\phi_d$ is highly non-gaussian. The sharp cut--off at $\phi_d \sim -11^{\rm o}$ is due to the interplay between $\epsilon_K$ and $\Delta M_{B_s}/\Delta M_{B_d}$ on one side and $S_{\psi K}$ and $B\to \rho\rho$ on the other (we remind the reader that the former extract $\beta^{\rm eff} = \beta + \phi_d$ and the latter $\alpha^{\rm eff} = \alpha - \phi_d$). The inclusion $V_{ub}$ into the fit lowers considerably the predicted value of $\sin (2 \beta)$ (as can be deduced from  Eq.~(\ref{sin2betaprediction}) and Fig.~\ref{fig:utsm}), thus impacting strongly the extraction of $\phi_d$. On the other hand, $\theta_A$ is essentially determined by the difference between the time dependent CP asymmetries in $B\to J/\psi K$ on one side and $B\to (\phi,\eta^\prime)K$ on the other; hence the outcome of the fit, Eq.~(\ref{mia1-thetaA-novub}), is largely independent on the inclusion of $V_{ub}$. Numerically we find that both the $\phi_d$ central value and error decrease by a factor of two: 
\bea
\phi_d &=& -\left(3.2^{+1.5}_{-1.3} \right)^{\rm o} \label{mia1-phid-vub}\\ 
\theta_A &=& -(3.5^{+2.2}_{-1.9} )^{\rm o} \label{mia1-thetaA-vub}
\eea
In this scenario we interpret the tension in the fit to the unitarity triangle in terms of NP contributions to the time dependent CP asymmetries in $b\to c \bar c s$ and $b\to s \bar s s$ modes. The discrepancy between the predicted and ``measured'' values of $\sin(2\beta)$ is explained by new physics contributions to $B_d$ mixing; the difference between $S_{\psi K}$ and $S_{(\phi,\eta^\prime) K}$ is induced by a new phase in the $b\to s$ penguin amplitude.

\subsection{Scenario II}
\label{subsec:mia2}
We now assume the absence of new physics contributions to $B_d$ mixing and investigate the possibility that the tension in the fit is induced by new effects in $K$ mixing. The discrepancy between the time dependent CP asymmetries in the $b\to c \bar c s$ and $b\to s$ penguin modes still requires an independent NP phase. We adopt the following parametrization:
\bea
\varepsilon_K & = & C_\varepsilon \; \left( \varepsilon_K \right)_{\rm SM} \; , \\
{\cal A}_{s\bar s s} & = &\left( {\cal A}_{s\bar s s} \right)_{\rm SM} \; e^{ i \theta_A} \;. 
\eea
Note that the expressions for the time--dependent CP asymmetries become:
\bea
S_{c\bar c s} & = & \sin \left[ 2 \beta \right] \; , \\
S_{s\bar s s} & = & \sin \left[ 2 (\beta + \theta_A) \right] \; .
\eea
In the right panel of Fig.~\ref{fig:mia} we show the 68\% C.L. (solid) and 95\% C.L. (dashed) allowed region in the $(C_\varepsilon,\theta_A)$ plane. The one-sigma ranges for these two parameters without the inclusion of $V_{ub}$ in the fit read: 
\bea
C_\varepsilon &=& (1.31 \pm 0.14)\; , \label{mia2-cepsilon-novub} \\
\theta_A &=& -(3.6 \pm 2.3)^{\rm o}  \; . \label{mia2-thetaA-novub}
\eea
The impact of $V_{ub}$ shifts only slightly these values:
\bea
C_\varepsilon &=& (1.28 \pm 0.13)  \; , \label{mia2-cepsilon-vub} \\
\theta_A &=& -(4.1 \pm 2.3)^{\rm o} \; .  \label{mia2-thetaA-vub}
\eea
In this scenario, the $\sin (2\beta)$ prediction coincides essentially with $S_{\psi K}$ and does not depend much on the inclusion of $V_{ub}$; hence the amount of new physics required to bring $\varepsilon_K$ in agreement with the rest of the fit is quite insensitive to the $V_{ub}$ constraint. 

\section{Operator analysis of new physics in the fit to the UT} 
\label{sec:oa-a}
From our previous discussion it is clear that the tension in the fits of the unitarity triangle are related to the presence of new physics either in $B_d$ or $K$ mixing. The effective Hamiltonian that describes meson mixing ($B_d$, $B_s$ and $K$) can be written as:
\bea
{\cal H}_{\rm eff} = \frac{G_F^2 m_W^2}{16 \pi^2} \left( V_{tq}^{} V_{tq^\prime}^*\right)^2 \left( 
\sum_{i=1}^5 C_i  O_i  +  \sum_{i=1}^3 \tilde C_i  \tilde O_i \right) + {\rm h.c.}  ,
\label{heffmixing}
\eea
where we have $(q,q^\prime) = (bd), (bs), (sd)$ for $B_d$, $B_s$ and $K$ mixing. The operators are defined as follows:
\bea
O_1 &=&  \left( \bar q^\prime_L \gamma_\mu q_L\right) \left( \bar q^\prime_L \gamma_\mu q_L\right) \label{q1}\\
O_2  &=&  \left( \bar q^\prime_R  q_L\right) \left( \bar q^\prime_R  q_L\right)  \\
O_3  &=&  \left( \bar q^{\prime\alpha}_R  q_L^\beta \right) \left( \bar q^{\prime\beta}_R  q_L^\alpha \right)  \\
O_4  &=&  \left( \bar q^\prime_R  q_L\right) \left( \bar q^\prime_L  q_R\right) \\
O_5  &=&  \left( \bar q^{\prime\alpha}_R  q_L^\beta \right) \left( \bar q^{\prime\beta}_L  q_R^\alpha \right) 
\eea
and $\tilde O_{1,2,3}$ are obtained from $O_{1,2,3}$ via the $L \leftrightarrow R$ substitution. In the following we will consider new physics contributions to the Wilson Coefficients of the operators $O_1$ and $O_4$ ($\tilde O_4$ has the same anomalous dimension and matrix element as $O_4$, hence we are really constraining $C_4 + \tilde C_4$). $O_1$ is the only operator that receives a non-negligible contribution in the SM. Contributions to $C_4$ are especially interesting because they are enhanced by QCD running effects and by chiral factors~\cite{Beall:1981ze} that appear in the calculation of their matrix elements; in particular, in the K mixing case these effects result in a two order of magnitude enhancement. We parametrize new physics contributions to the various Wilson coefficients as:
\bea
\delta C_{\rm 1,4}^{B_q,K} (\mu^0) = - \frac{1}{G_F^2 m_W^2} \frac{e^{i \varphi}}{\Lambda^2} \;.
\label{eq:wcnp}
\eea
where we retained a factor $1/(16\pi^2)$ to take into account a possible loop-suppression of NP effects\footnote{This factor appears in the deÞnition of the effective Hamiltonian~(\ref{heffmixing}).} and we decided to factor out the CKM couplings. The factor $-1$ is introduced because we know from the model independent analysis of Sec.~\ref{subsec:mia1} that the required NP phase has to be negative. Combining Eqs.~(\ref{heffmixing}) and (\ref{eq:wcnp}), the NP contribution to the effective Hamiltonian is 
\bea
\delta {\cal H}_{\rm eff} =  - \frac{ \left( V_{tq}^{} V_{tq^\prime}^*\right)^2}{16 \pi^2} \;
\frac{e^{i \varphi}}{\Lambda^2}  \; O_{i} + {\rm h.c.}
\eea
and the scale $\Lambda$ absorbs every NP coupling, mass scale and loop function apart from the CKM matrix and the typical $1/(16 \pi^2)$. Using Eq.~(\ref{eq:wcnp}) we obtain $C_{\rm NP}/C_{\rm SM} \simeq - e^{i \varphi} (700\; \gev/\Lambda)^2$; hence for $\Lambda \sim 700 \; \gev$ the new physics and SM contributions to the Wilson coefficients are of similar size. In the remainder of this section we show the bounds on $\Lambda$ that we obtain in the two scenarios we introduced in Sec.~\ref{sec:mia}. We will also consider the possibility of simultaneous NP contributions to $(B_d , B_s)$ and $(B_d, B_s, K)$ mixing
\subsection{New physics in $B_d$ mixing}
\label{subsec:oa-d}
We begin by assuming that new physics contributes to a single operator relevant to $B_d$ mixing (this corresponds to Scenario I of Sec.~\ref{subsec:mia1}) and take $\delta^{\rm NP} C_1^{B_d} \neq 0$. We do not consider NP contributions to the other $(\bar b d)(\bar b d)$ operators that appear in Eq.~(\ref{heffmixing}) because their hadronic matrix elements are all very similar (e.g. no large chiral enhancement of LR operators) and the outcome of the analysis does not change appreciably. 
Note that we introduce new physics to the $B_d$ mixing amplitude only; as a consequence, the phase $\phi_d$ is non vanishing and, since we do not allow contributions to $B_s$ mixing, the ratio $\Delta M_{B_s} /\Delta M_{B_d}$ will be affected too. Using the expression Eq.~(\ref{eq:wcnp}) for the generic NP contribution to $C_1^{B_d}$ we have:
\FIGURE[p]{
\epsfig{file=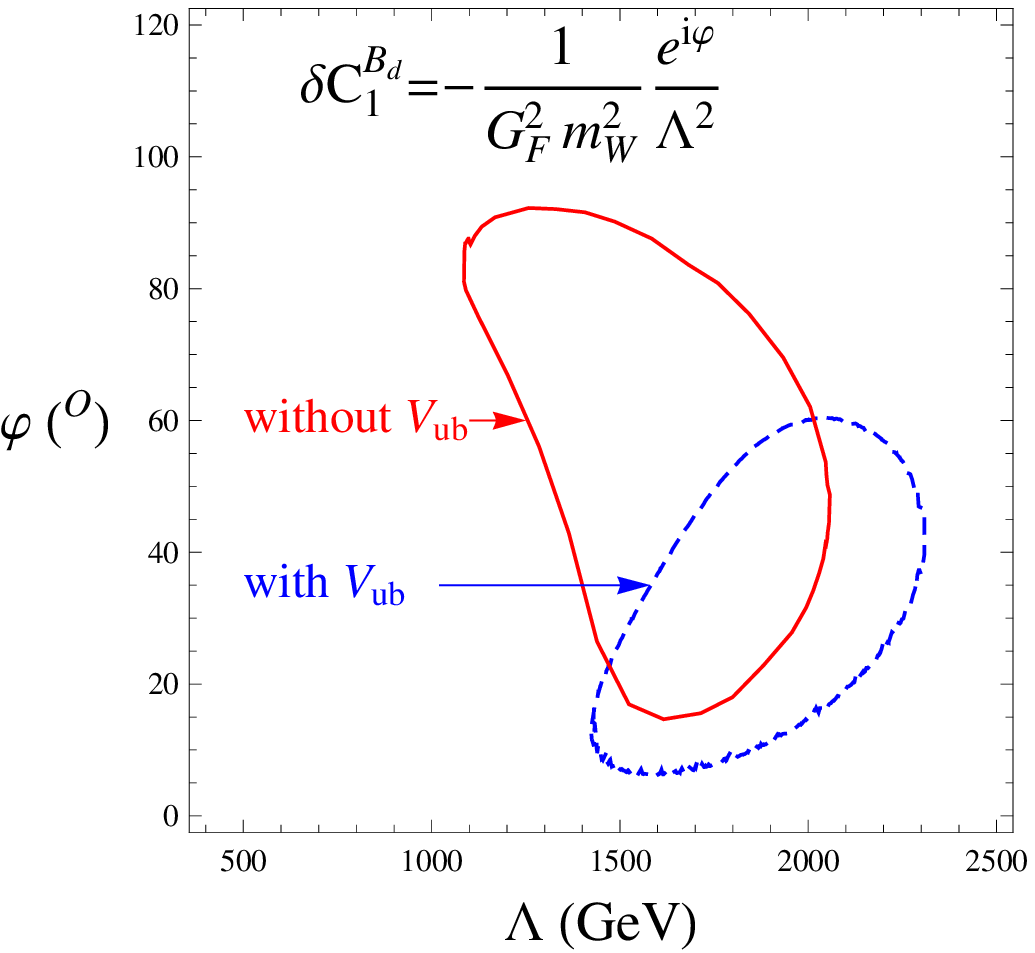,width= 0.45\linewidth}
\caption{New physics in $B_d$ mixing. \label{fig:oa-d}}}
\bea
\hskip -1cm
M_{12}^d & = & \left| M_{12}^d \right|_{\rm SM} \; e^{2 i \beta} \; \left( 1 - \frac{e^{i \varphi}}{\kappa \Lambda^2} \right)
\label{kappa}
\eea
where $\kappa = G_F^2 m_W^2 S_0 (x_t)$. From the model independent results we obtained in Sec.~\ref{sec:mia}, we see that the phase $\phi_d =1/2  \arg ( 1 - e^{i\varphi}/\kappa \Lambda^2)$ is negative, implying $\varphi > 0$. In Fig.~\ref{fig:oa-d} we show the allowed regions in the $\Lambda,\varphi$ plane. We write a chi-squared that contain all the observables we discussed in Secs~\ref{sec:sm} and \ref{sec:mia} (namely $\varepsilon_K$, $\Delta M_{B_{d,s}}$, $V_{cb}$, $|V_{ub}|$, $\gamma$ from $D^{(*)} K^{(*)}$ decays, $\alpha$ from $B\to \rho\rho$, and $S_{\psi K_S}$) and minimize with respect to all variables (including $\bar \rho$ and $\bar \eta$). The contours are such that their projections on the axis correspond to the one--dimensional 68\% C.L. regions for $\Lambda$ and $\varphi$. The green (dashed) and blue (solid) contours are obtained with and without the inclusion of $|V_{ub}|$ in the fit, respectively. The presence of the upper limit $\Lambda \lesssim  2.3 \; {\rm TeV}$ reflects the two sigma effects Eqs.~(\ref{mia1-phid-novub}) and (\ref{mia1-phid-vub}). The lower bound $\Lambda \gtrsim 1. \; {\rm TeV}$ is a direct consequence of NP contributions to $X_{sd}$:
\bea 
\frac{\Delta M_{B_s}}{\Delta M_{B_d}} & = & \left( \frac{\Delta M_{B_s}}{\Delta M_{B_d}}  \right)_{\rm SM} \; \left| 1 -  \frac{e^{i \varphi}}{\kappa \Lambda^2}  \right|^{-1} \; .
\eea
The qualitative impact of the $V_{ub}$ constraint can be inferred from the analysis of Sec.~\ref{subsec:mia1} and from Fig.~\ref{fig:mia}. A reduction in the absolute size of $\phi_d$ translates into smaller values for $\varphi-\pi$. Finally, it is interesting to extract the predicted value for $\sin (2\beta)$:
\bea
\sin (2\beta) = \cases{ 0.82 \pm 0.10 & without $V_{ub}$ \cr 0.73 \pm 0.03 & with $V_{ub}$\cr } \; .
\eea
The comparison of this result with the SM prediction given in Eq.~(\ref{sin2betaprediction}), shows that in this scenario the tension between $\sin (2\beta)$ and the CP asymmetries in the $\phi K$ and $\eta^\prime K$ channels
is somewhat eased: $S_{(\phi+\eta^\prime)K}$ deviates from $\sin (2\beta)$ at the 2.2/2.1 $\sigma$ level with/without the inclusion of $V_{ub}$.

\subsection{New physics in both $B_d$ and $B_s$ mixing} 
\label{subsec:oa-dsk}
In this section we modify the analysis of Sec.~\ref{subsec:oa-d} by allowing simultaneous identical NP contributions to $B_d$ and $B_s$ mixing: $\delta C_1^{B_d} = \delta C_1^{B_s}$. This approach is inspired by a Minimal Flavor Violating (MFV) ansatz\footnote[2]{However, we stress that if the $B$-$CP$ anomalies we discussed in here are confirmed, that would be quite inconsistent with the general notions and expectations of models based on MFV.} in which NP contributions to $M_{12}^{d,s}$ are identical up to CKM factors:
\bea 
\delta M_{12}^d \propto - \left( V_{tb}^{} V_{td}^{*} \right)^2 \frac{e^{i \delta}}{\Lambda^2} 
\quad \quad {\rm and} \quad \quad
\delta M_{12}^s \propto -  \left(  V_{tb}^{} V_{ts}^{*} \right)^2 \frac{e^{i \delta}}{\Lambda^2} 
\;.
\eea
\FIGURE[p]{
\epsfig{file=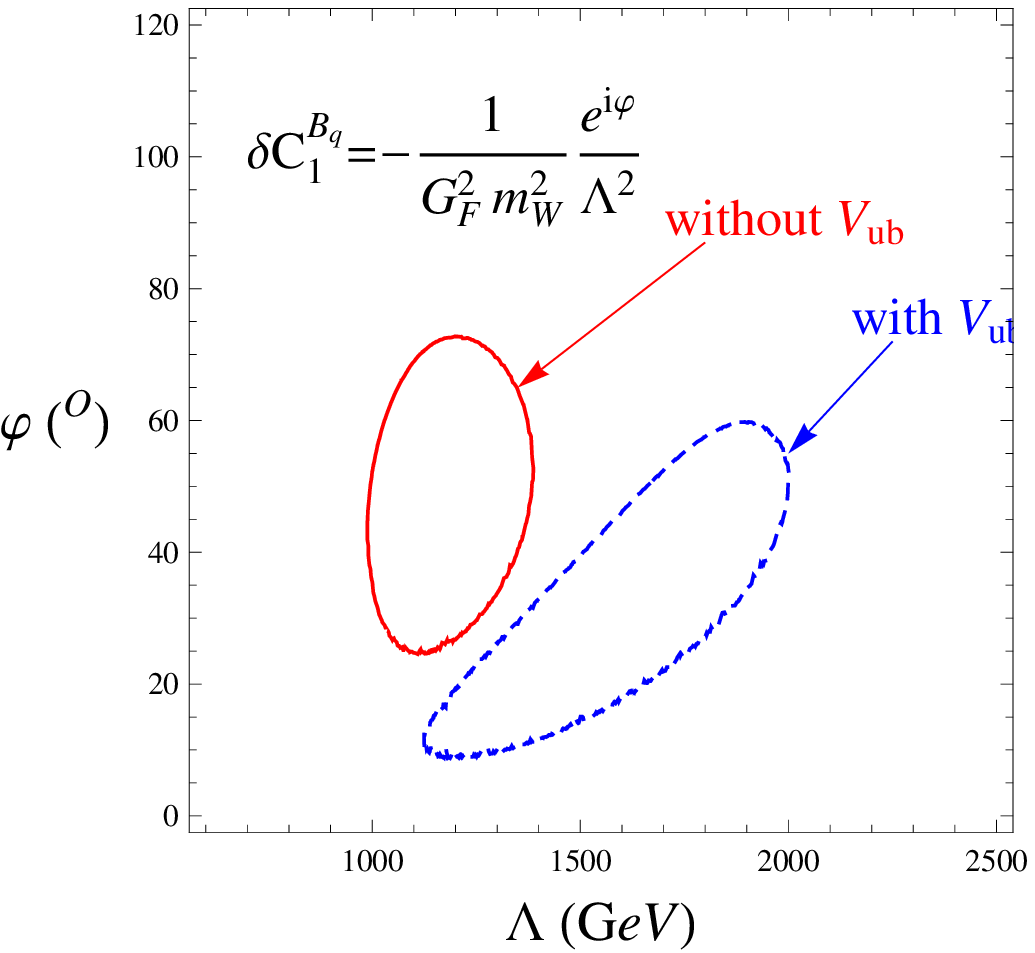,width= 0.45\linewidth}
\caption{New physics contributions to $B_d$ and $B_s$ mixing. \label{fig:oa-ds}}
}
\noindent 
It is important to stress that we introduce this complex contribution only in flavor changing operators involving quarks. In explicit NP models that implement this idea, e.g. extra $Z^\prime$ with flavor changing quark couplings, it is important to keep under control phases that appear in interactions that conserve flavor or involve leptons (the latter, in particular, are constrained by CP asymmetries in exclusive $b\to s\ell^+\ell^-$ decays). The inclusion of a complex correction to $M_{12}^s$ is of great interest because it allows to reconcile the constraint coming from the time dependent CP asymmetry in the $B\to J/\psi \phi$ system with the rest of the fit. In the numerics we utilize the HFAG combination of D0 and CDF data: $\phi_s = -0.76^{+0.37}_{-0.33}$ or $\phi_s = -2.37^{+0.33}_{-0.37}$~\cite{Barberio:2008fa} (information from the flavor specific life time and the semileptonic asymmetry is included).

In this scenario the mass differences $\Delta M_{B_s}$, $\Delta M_{B_d}$, as well as the $B_q$ mixing phases are affected but the ratio $\Delta M_{B_s} /\Delta M_{B_d}$ does not receive contributions. This is because the relative impact of NP on the $B_q$ systems is identical. The measurements that we include into the fit are $\varepsilon_K$, $\Delta M_{B_{d,s}}$, $V_{cb}$, $|V_{ub}|$, $\gamma$ from $D^{(*)} K^{(*)}$ decays, $\alpha$ from $B\to \rho\rho$, $S_{\psi K_S}$ and $S_{J/\psi \phi}$. The result of the analysis is summarized in Fig.~\ref{fig:oa-ds}. A striking feature of these results is the large impact that the inclusion of $V_{ub}$ has on the allowed regions in the $(\Lambda,\varphi)$ plane. The impact of $V_{ub}$ is to require a smaller $\phi_d$ phase because the discrepancy between the direct and indirect determinations of $\sin (2\beta)$ decreases in absolute value. On the other hand the observed discrepancy in the phase measured in the $J/\psi \; \phi$ system still points to large effects. The friction between these two competing effects results is responsible for the large shifts in the contours obtained with and without $V_{ub}$. The predictions that we obtain for $\sin (2\beta)$ are:
\bea
\sin (2\beta) = \cases{ 0.88 \pm 0.07 & without $V_{ub}$ \cr 0.75 \pm 0.03 & with $V_{ub}$\cr } \; .
\eea
In this case, the tension between the global extraction of $\sin(2\beta)$ and the $(\phi,\eta^\prime)K$ CP asymmetries is unaffected.
\subsection{New physics in $K$ mixing}
\label{subsec:oa-k}
The scenario described in Sec.~\ref{subsec:mia2} corresponds to new physics contributions to the $K$ mixing amplitude only. We implement this framework by allowing contributions to either $C_1^K$ or $C_4^K$: all new physics effects are confined to $\varepsilon_K$. Note that this time  we consider separately possible NP effects in the LR operator $O_4^K$: because of the large QCD running effects on its Wilson coefficient and of the chiral enhancement of its matrix element, the bounds that we extract for this case are about one order of magnitude stronger then the ones we obtain for new physics in  $O_1$. The explicit formula for $\varepsilon_K$ that we use to study NP contributions to the Wilson coefficients of the operators $O_1^K$ and $O_4^K$ is obtained from the SM formula (\ref{ek}) via the substitution:
\bea
S_0 (x_t) & \rightarrow & S_0 (x_t) \left[ 1 - \frac{e^{i \varphi}}{\kappa \Lambda^2} \right]  ,
\;\; \mbox{for NP in } C_1^K  \\
S_0 (x_t) & \rightarrow & S_0 (x_t) \Big[ 1 - \frac{e^{i \varphi}}{\kappa \Lambda^2} 
\underbrace{\frac{B_4  K_{44}}{\hat B_K \eta_2} \;
   \frac{3 \; m_K^2}{4 (m_s + m_d)^2}}_{\equiv \; \chi} \Big] ,
\;\; \mbox{for NP in } C_4^K
\label{ek-o4}
\eea
where $\kappa = G_F^2 m_W^2 S_0 (x_t)$, $m_s =  (100 \pm 10)\; \mev$~\cite{Bernard:2007ps,Allton:2008pn}, $m_d = (4.6 \pm 3) \; \mev$~\cite{Bernard:2007ps}, $B_4 = 1.03 \pm 0.06$~\cite{Ciuchini:1998ix} and $K_{44} \simeq 3.7$ is calculated below. The factor $\chi$, whose numerical estimate is given below in Eq.~(\ref{eq:chi}), quantifies the different impact that new physics contributions to the Wilson coefficients $C_1$ and $C_4$ have on $\varepsilon_K$ and is enhanced by chiral and QCD--running effects. The masses of the strange and down quarks are defined in the $\overline{MS}$ scheme at the scale $\mu_L \sim 2 \; \gev$ and are taken from Ref.~\cite{Bernard:2007ps,Allton:2008pn} (the actual value of $m_s$ that we adopt reflects the dispersion of several lattice results). $B_4 = B_4^{\overline{MS}} (\mu_L)$ is the bag parameter of the operator $O_4$ and it has been calculated in quenched lattice QCD; the value we use is taken from Ref.~\cite{Ciuchini:1998ix}. Note that we have:
\bea
\frac{\langle O_4^K (\mu_L)  \rangle}{\langle O_1^K (\mu_L)  \rangle} & = &
 \frac{3 \; m_K^2}{4 (m_s + m_d)^2} \; \frac{B_4}{ B_K}  \; ,
\eea
Finally, QCD effects in the running of the Wilson coefficients between $\mu_H$  and $\mu_L$ are summarized in the matrix $K_{rs}$:
\bea
K_{rs} & = & \sum_i \left( b_i^{(r,s)} + \eta  \; c_i^{(r,s)} \right) \eta^{a_i}  \; ,\\
C_r (\mu_L) & = & \sum_s K_{rs} \; C_s (\mu_H) \; ,
\eea
where $\eta = \alpha_s (M_H)/\alpha_s (m_t)$ and the magic numbers $a_i$, $b_i$ and $c_i$ have been calculated for $\mu_L = 2 \; \gev$ in Ref.~\cite{Ciuchini:1998ix}. In this analysis we take $\mu_H = m_t$. The MS-bar scheme dependence of the bag parameters $B_i$ can be removed by including part of the QCD running effects into their definition, thus leading to the introduction of the hat parameters $\hat B_i$. Eq.~(\ref{ek-o4}) is justified because we have $B_K K_{11} = \hat B_K \eta_2$. Putting everything together we find
\bea
\chi = (157 \pm 33) \left( \frac{0.720}{\hat B_K} \right) \left( \frac{0.5765}{\eta_2} \right) . 
\label{eq:chi}
\eea
The presence of NP in $M_{12}^K$ affects the extraction of $\sin (2\beta)$ from $S_{\psi K}$. In the present case NP contributions to K mixing are proportional to $(V_{td} V_{ts}^*)^2$; hence their effect is $O(1)$ on $\varepsilon_K \sim {\rm Im} M_{12}^K$ but only $O(0.1\%)$ on ${\rm Re} M_{12}^K$. In the following we will neglect such corrections to $S_{\psi K}$. In Fig.~\ref{fig:oa-k} we show the results of the analysis. The projections of the blue (solid) and green (dashed) contours onto the $\Lambda$ and $\varphi$ axes correspond to the one--dimensional 68\% C.L. ranges. 
\FIGURE[t]{
\epsfig{file=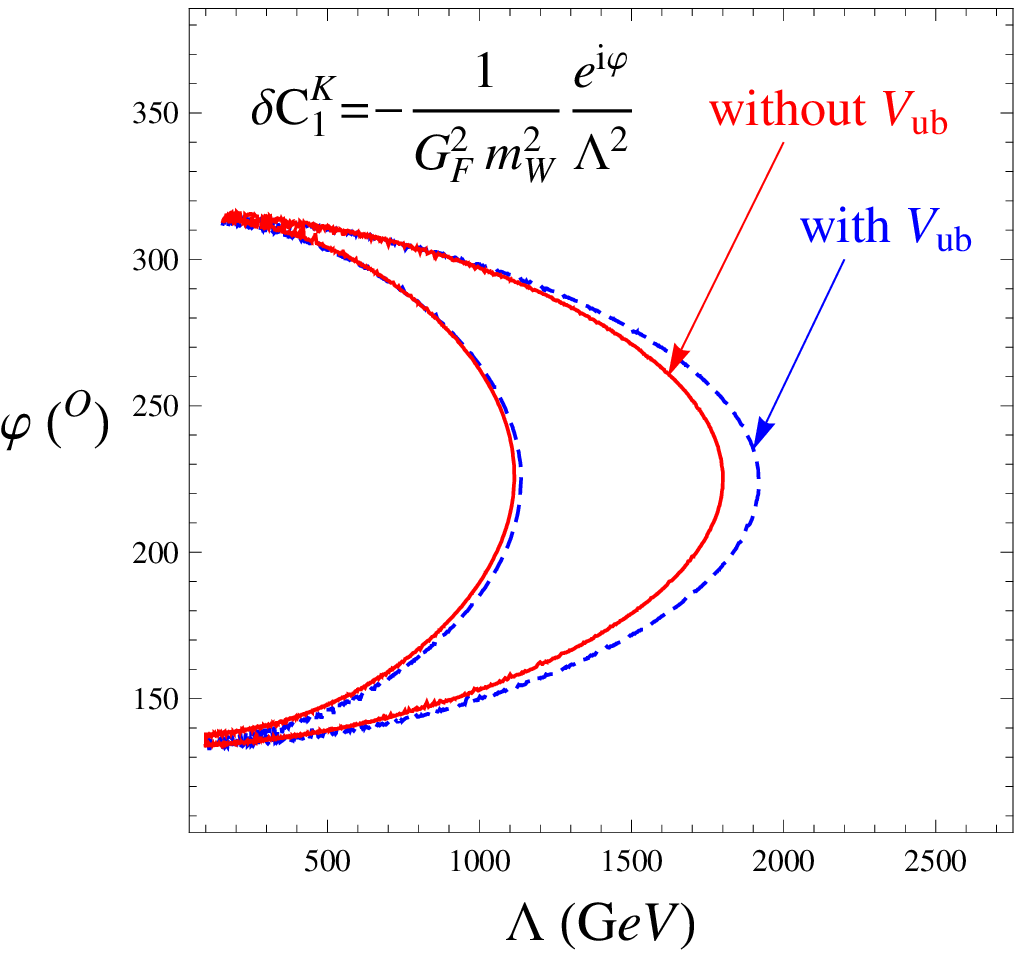,width= 0.45\linewidth} 
\epsfig{file=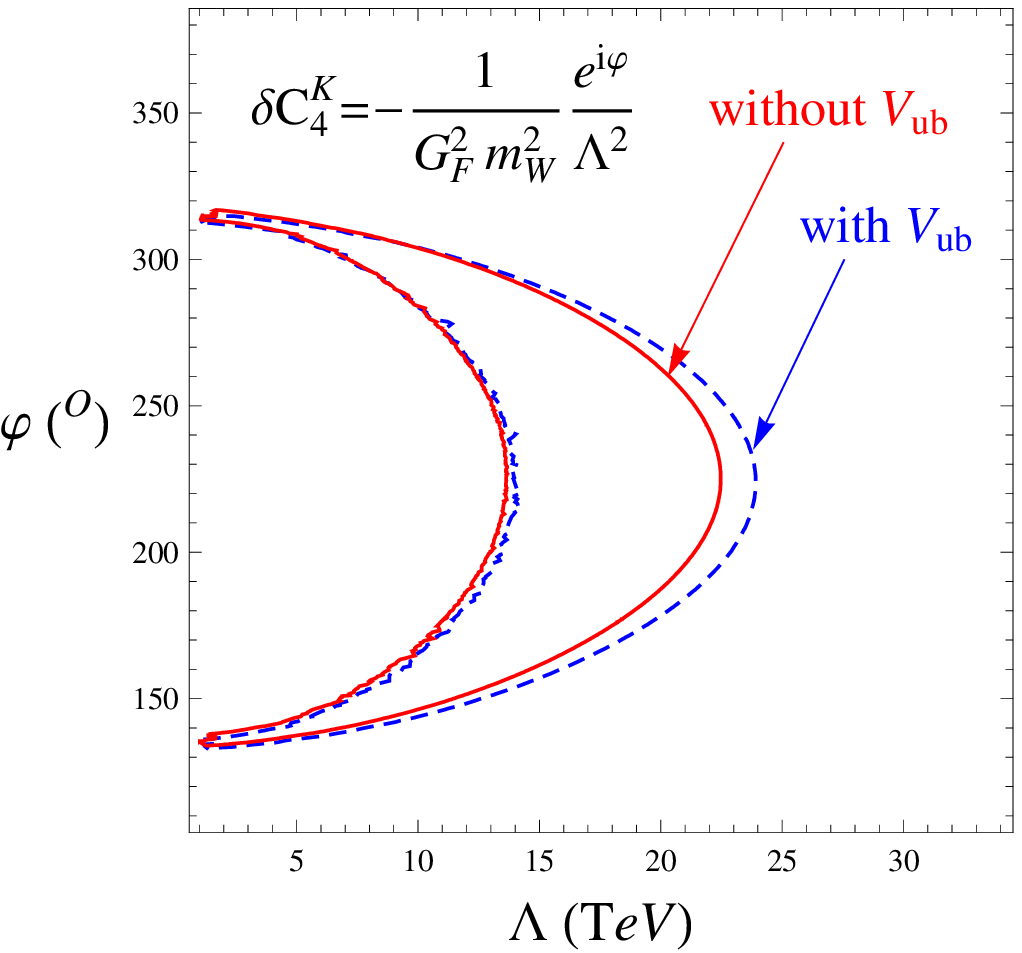,width= 0.45\linewidth} 
\caption{New physics in $K$ mixing. \label{fig:oa-k}}}

Note that the inclusion of $V_{ub}$ in the fit does not impact appreciably the allowed regions in the $(\Lambda,\varphi)$ plane. In this scenario the role of NP effects is to reconcile the $\varepsilon_K$ constraint with the rest of the fit ($V_{cb}$, $V_{ub}$, $\Delta M_{B_q}$, $\alpha$, $\gamma$ and, especially, $S_{\psi K}$). In particular, the inclusion of $S_{\psi K}$ in the fit renders the latter quite insensitive to $V_{ub}$. In general $V_{ub}$ tends to slightly improve the overall consistency of the fit to the UT within the SM, therefore after the inclusion of $V_{ub}$ the bounds on $\Lambda$ become slightly weaker. 

Another interesting feature of Fig.~\ref{fig:oa-k} is the absence of a lower limit on the scale of NP. This happens because $\varepsilon_K$ is given by the imaginary part of $M_{12}^K$; hence a given correction can be obtained for any arbitrarily small $\Lambda$ by appropriately choosing $\varphi$. Note that the NP contribution to $M_{12}^K$ is proportional to $-e^{i (2 \beta +\varphi)}/\Lambda^2$ and in the limit $\varphi \to (\pi-2 \beta,2 \pi - 2 \beta)$ the correction induced on $\varepsilon_K$ vanishes: this feature is evident in Fig.~\ref{fig:oa-k} in which the asymptotic values of $\varphi$ in the limit $\Lambda \to 0$ are very close to $\pi-2 \beta$ and $2 \pi -2 \beta$. 

The upper bounds on the NP scale that we extract  are about $1.8 \div 1.9 \; {\rm TeV}$ and $23 \div 24 \; {\rm TeV}$ for the $C_1^K$ and $C_4^K$ scenarios, respectively. 

\section{Operator analysis of new physics in $b\to s$ amplitudes}
\label{sec:oa-b}
In this section we interpret the difference between the time dependent CP asymmetries $S_{\psi K}$ and $S_{\phi,\eta^\prime}$ in terms of new physics contributions to the QCD or EW penguin operators. The effective Hamiltonian responsible for the $B\to (\phi,\eta^\prime) K_S$ amplitudes is:
\bea
{\cal H}_{\rm eff} =  \frac{4 G_F}{\sqrt{2}} V_{cb}^{} V_{cs}^{*} \left(
\sum_{i=1}^{6} C_i (\mu) O_i (\mu) + \sum_{i=3}^{6} C_{iQ} (\mu) O_i (\mu) \right) 
+{\rm h.c.} \; .
\eea
The definition of the various operators can be found, for instance, in Ref.~\cite{Huber:2005ig}. Here we focus on two operators whose matching conditions are are likely to receive new physics contributions: 
\bea
Q_4 & = & \left( \bar s_L \gamma^\mu T^a b_L \right) \sum_q \left(\bar q \gamma_\mu T^a q\right)  \; .\\
Q_{3Q} & = & \left( \bar s_L \gamma^\mu  b_L \right) \sum_q Q_q \left( \bar q \gamma_\mu q \right) \; .
\eea
\FIGURE[t]{
\epsfig{file=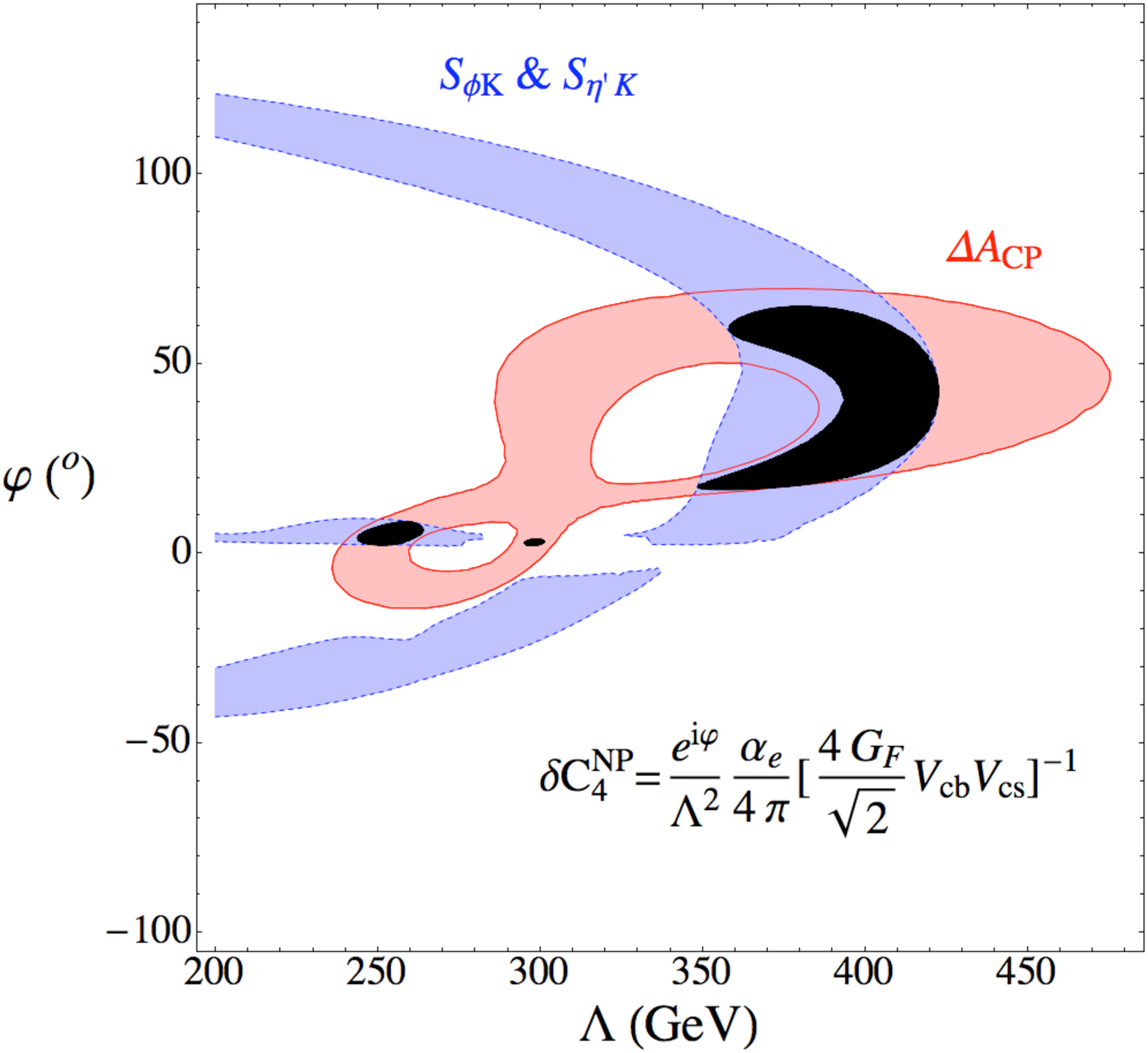,width= 0.45\linewidth}
\epsfig{file=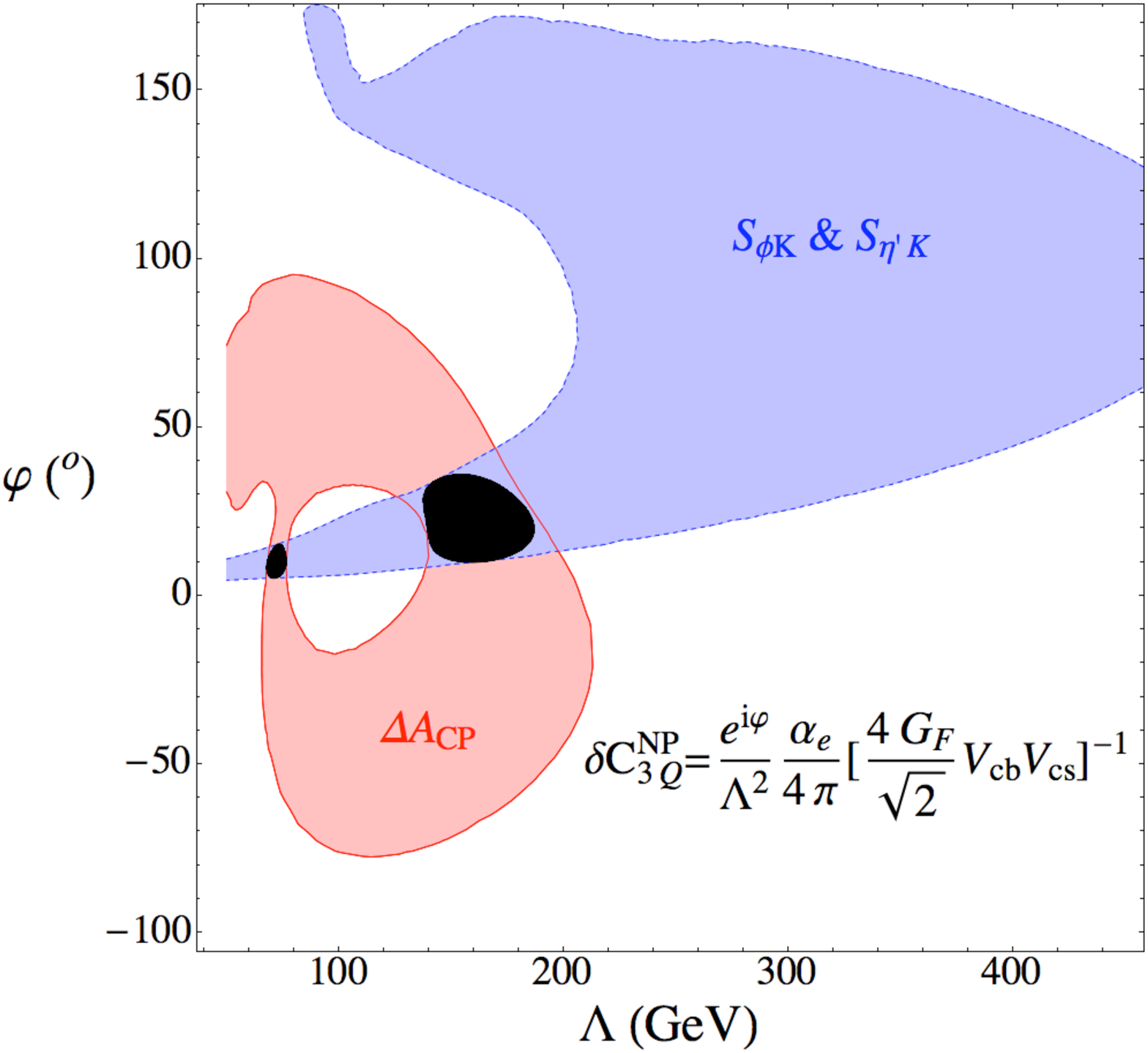,width= 0.45\linewidth}
\caption{New physics in $b\to s$ penguin amplitudes. \label{fig:thetaA}}}
We adopt the following parametrization of new physics effects:
\bea
\delta C_{4,3Q} = \frac{e^{i \varphi}}{\Lambda^2} \frac{\alpha_{s,e}}{4\pi} \left[ \frac{4 G_F}{\sqrt{2}} V_{cb}^{} V_{cs}^* \right]^{-1} \; ,
\eea
where we kept the coupling and loop suppression typical of QCD and EW penguins (the factor $\alpha_{s,e}/(4\pi) = (e^2,g_s^2)/(16 \pi^2)$). Note that we have absorbed all new physics couplings in the effective scale $\Lambda$. It is important to notice that once we introduce new physics contributions in these penguin operators we induce important effects in the $B\to K \pi$ system as well; in particular, we will consider the difference of CP asymmetries $\Delta A_{CP} = A_{CP} (B^-\to K^- \pi^0)- A_{CP} (\bar B^0 \to K^- \pi^+)$. In order to describe the impact of the new physics coefficients onto the $\phi K$, $\eta^\prime K$ and $K\pi$ system, we follow the QCD factorization analysis of Refs.~\cite{Beneke:2003zv,Beneke:2002jn,Beneke:2005pu}. Within the SM we find:
\bea
\delta S_{\phi K} & = & 0.03 \pm 0.01 \; , \label{aphik-qf} \\
\delta S_{\eta^\prime K} & = & 0.01 \pm 0.025 \; , \label{aetapk-qf} \\
\Delta A_{CP}^{\rm SM} &=& (2.2 \pm 2.4) \%  \; . \label{dacp-qf}
\eea
The errors in Eqs.~(\ref{aphik-qf}--\ref{dacp-qf}) are obtained by varying simultaneously all the hadronic inputs that we take from Refs.~\cite{Beneke:2003zv,Beneke:2002jn,Beneke:2005pu}. For what concerns the $B\to K\pi$ asymmetries we quote also the results that we obtain for the various topological amplitudes: $C/T = (0.16\pm 0.08) - i (0.08\pm 0.05)$, $P_{EW}/T = -(0.67 \pm 0.03) - i (0.004\pm 0.06)$, $|T/P| = 0.24 \pm 0.03$. $T$ and $C$ correspond to the color allowed and color suppressed matrix elements of the tree-level operators $O_{1,2}$; $P$ and $P_{EW}$ are the one loop matrix elements of the QCD ($O_{3-6}$) and EW ($O_{7-10}$) penguin operators, respectively. In our conventions, the definitions of these amplitudes include the magnitude of the corresponding CKM factors but not their phases. 

In the left and right panels of Fig.~\ref{fig:thetaA}, we present the allowed regions of the $(\Lambda,\varphi)$ plane in presence of new physics contributions to $C_4$ and $C_{3Q}$. The blue and red shaded regions are obtained using the constraints from $S_{(\phi,\eta^\prime) K}$ and $\Delta A_{CP}$, respectively. The black areas are obtained by requiring both constraints simultaneously. The excluded regions within the $\Delta A_{CP}$ contours correspond to a part of the parameter space that yields a too large value of $\Delta A_{CP}$. The irregular behavior of the $S_{(\phi,\eta^\prime) K}$ contours is due to the complicated structure of the theoretical errors on these quantities. For each point in the $(\Lambda,\varphi)$ plane we determine the theoretical error by varying all the hadronic inputs; the resulting two--dimensional error function is then utilized in the chi--squared fit. 

The most important result of this analysis is the existence of an upper limit on the effective scale of about 400 GeV (200 GeV) for new physics contributions to QCD (EW) penguin operators. 
\section{Summary}
\label{sec:discussion}
We discussed several anomalies involving CP asymmetries in $B$ and $B_s$ decays. The measured CP asymmetry in $B\to J/\psi K_S$ when compared with the SM prediction from the fits of the UT seems to be too small by about $15\%$ and this hints to new physics in either $B_d$ or $K$ mixing. The non-vanishing differences between the time dependent CP asymmetries in $B\to J/\psi K_S$ and $B\to (\phi,\eta^\prime) K$ modes monitors the presence of new physics in $b \to s$ transitions. The latter is also hinted at by the direct CP asymmetries in the $K\pi$ system ($\Delta A_{CP}$). The large asymmetry in $B_s \to J/\psi \phi$ also indicates a non-vanishing beyond the SM phase in $B_s$ mixing on the face of a negligible asymmetry in the SM. 

The tension in the fit to the UT can, for example, be explained with an extra phase in $M_{12}^d$ whose value is found to be $\phi_d \approx -(3 \pm 1.5)^{\rm o}$ (we obtain $-(9 \pm 3)^{\rm o}$ if no use of $V_{ub}$ is made) or by new physics in $\varepsilon_K$ for which we find $C_\varepsilon \approx 1.3 \pm 0.1$, where in the SM, $C_\varepsilon = 1$. The anomaly in the $(\phi,\eta^\prime)$ system points to a new phase in $b\to s$ amplitudes for which we obtain $\theta_A \approx -(4 \pm 2)^{\rm o}$.

These results can be interpreted in an effective Hamiltonian formalism in terms of NP contributions to Wilson coefficients. In this way we can translate these hints for NP into scales at which we expect to find accelerator signals. Our results for different physics scenarios are summarized in the following table:
\begin{center}
\begin{tabular}{|c|c|c|c|} \hline
Scenario  & Operator & $\Lambda\; ({\rm TeV})$ & $\varphi \;({}^{\rm o})$ \\ \hline 
\vphantom{$\cases{a \cr b\cr x \cr }$} 
$B_d$ mixing & $O_1^{(d)}$ & 
$\cases{1.1 \div 2.1 & no $V_{ub}$ \cr 1.4 \div 2.3 & with  $V_{ub}$ }$ & 
$\cases{15 \div 92 & no $V_{ub}$ \cr 6 \div 60 & with  $V_{ub}$ }$ \\  \hline
\vphantom{$\cases{a \cr b\cr x \cr }$} 
$B_d = B_s$ mixing & $O_1^{(d)} \; \& \; O_1^{(s)}$ & 
$\cases{1.0 \div 1.4 & no $V_{ub}$ \cr 1.1 \div 2.0 & with  $V_{ub}$ }$ & 
$\cases{25 \div 73 & no $V_{ub}$ \cr 9 \div 60 & with  $V_{ub}$ }$ \\  \hline
\vphantom{$\cases{a \cr b\cr x \cr }$} 
$K$ mixing & 
$\begin{array}{l} 
O_1^{(K)} \\ 
O_4^{(K)} \\ 
\end{array}$ & 
$\begin{array}{l} 
< 1.9 \\  
< 24 \\ 
\end{array}$ & 
$130\div 320$ \\ \hline
\vphantom{$\cases{a \cr b\cr x \cr }$} 
${\cal A}_{b\to s}$ & 
$\begin{array}{l} 
O_{4}^{b\to s} \\  
O_{3Q}^{b\to s} \\ 
\end{array}$ & 
$\begin{array}{l} 
.25 \div .43 \\  
.09 \div .2  \\ 
\end{array}$ & 
$\begin{array}{l} 
0 \div 70 \\  
0 \div 30 \\ 
\end{array}$ \\ 
\hline
\end{tabular}
\end{center}
Our main finding is that, no matter what kind of new physics is invoked to explain these effects, its effective scale is bounded from above at few TeV. The only exception are NP contributions solely confined to the LR operator in $K$-mixing; however, it should be stressed that if NP affects only $K$ mixing, then it cannot explain the difference in the extraction of $sin 2 \beta$ from $B\to J/\psi K_S$ and $B\to (\phi , \eta^\prime)K$ and, in addition, it cannot account for both $\Delta A_{CP} (k \pi)$ and the asymmetry in $B_s \to J/\psi \phi$. 

Finally, let us comment on a similar analysis for the scale of NP  presented in Ref.~\cite{Bona:2007vi}. One  important difference   concerns the treatment of $\varepsilon_K$: we utilize the  recent (2+1)-flavors determination of $\hat B_K$ from the RBC collaboration and the estimation of the effect of the 0--isospin $K\to \pi \pi$ amplitude on $K$ mixing (the factor $\kappa_\varepsilon$)\cite{Buras:2008nn,Buras:2009pj,Anikeev:2001rk,Andriyash:2003ym,Andriyash:2005ax}. The combined effect of the updated values for these parameters is to strengthen the impact of the $\varepsilon_K$ constraint on the fit to the UT and to introduce the $\gtrsim 2 \sigma$ discrepancy responsible for the upper limits on the NP scales that we find. Another important difference  of the present analysis from Ref.~\cite{Bona:2007vi} is in the  treatment of NP effects on the CP asymmetries in $b\to s$ penguin and $K \pi$ modes, both of which are included in our analysis and not in Ref.~\cite{Bona:2007vi}.

Lastly, we want to briefly comment on the possibility of resolving these anomalies within the SM. In particular the impact of indirect CP violation in the $K$ system, $\epsilon_K$, depends crucially on the hadronic matrix elements $\hat B_K$ and on the precise value of $|V_{cb}|$ (we remind the reader that the $\rho$ and $\eta$ dependent part of $\varepsilon_K$ is proportional to $|V_{cb}|^4$). While a rather large shift in a single input parameter (such as $\hat B_K$ or $V_{cb}$) is needed to reduce the discrepancy between the fitted and measured (via $B\to J/\psi K_S$) values of $\sin 2\beta$, a correlated set of smallish  shifts in several inputs, while implausible, can certainly not be ruled out. However, for the effects that we discuss to disappear, any such problems in the hadronic matrix elements from the lattice and/or elsewhere will only suffice, {\it if} simultaneously it is proven that the  $B_s \to J/\psi \phi$ asymmetry and the smaller values of $\sin 2 \beta$ from penguin modes were all a statistical fluctuation. 

Finally, let us summarize the impact of future experimental and theoretical progress on the anomalies we considered in this analysis. The most promising developments will be the high precision measurement of the $B_s \to J/\psi \phi$ asymmetry, additional lattice-QCD calculations of $\hat B_K$, the inclusion of $O(\alpha_s^2)$ and $O(\alpha_s/m_b)$ corrections in the global fit to inclusive $b\to c\ell \nu$ decays for the extraction of $|V_{cb}|$, and the calculation of the parameter $\kappa_\varepsilon$ introduced in (\ref{ek}) using 2+1 flavors lattice-QCD ($\kappa_\varepsilon$ is controlled by the matrix element of the QCD penguin operator $O_6$ - not presently calculable with good accuracy with lattice QCD methods~\cite{Li:2008kc} - and can be extracted, within the SM, from the measurement of $\varepsilon^\prime/\varepsilon$ and the lattice determination of the matrix element of the electro-weak penguin operator $O_8$.). Because of recent progress in lattice calculations, the errors on $\hat B_K$ and $\kappa_\varepsilon$ do not impact too strongly the fit to the unitarity triangle anymore; therefore new determinations of these parameters will serve mainly to build confidence in the central values that we are presently using. On the other hand, the role of $|V_{cb}|$ is of the utmost importance: if the inclusion of higher order QCD corrections to inclusive semileptonic B decays will help closing the $2\sigma$ gap between inclusive and exclusive determinations of $|V_{cb}|$, the discrepancies we considered in this work will be strongly reinforced. Improved determinations of CP asymmetries in $B \to K\pi$ and $b\to s$ penguin modes will most probably have to wait for LHC-b and/or super-B factories~\cite{Hashimoto:2004sm,Bona:2007qt}; unfortunately the interpretation of these discrepancies relies heavily on QCD-factorization methods and suffers from our lack of control over power corrections.

\smallskip

\section{Acknowledgements}

We want to thank Kaustubh Agashe, Marcella Bona, Andrezj Buras, Hai-Yang Cheng, Diego Guadagnoli, Mikihiko Nakao and Viola Sordini for discussions. This research was supported in part by the U.S. DOE contract No.DE-AC02-98CH10886(BNL).

\end{document}